\documentclass[apjl]{emulateapj}
\usepackage{natbib,amssymb,amsmath,graphicx}
\shorttitle{Tracing Electron Beams in the Sun's Corona}
\shortauthors{Chen et al.}

\begin{document}

\title{Tracing Electron Beams in the Sun's Corona with \\ Radio Dynamic Imaging Spectroscopy}

\author{Bin Chen}
\affil{Department of Astronomy, University of Virginia, Charlottesville, VA 22904 USA}

\author{T. S. Bastian}
\affil{National Radio Astronomy Observatory, Charlottesville, VA 22903, USA}

\author{S. M. White}
\affil{Air Force Research Laboratory, Kirtland Air Force Base, New Mexico, USA}

\author{D. E. Gary}
\affil{Center for Solar-Terrestrial Research, New Jersey Institute of Technology, Newark, NJ 07102, USA}

\author{R. Perley}
\affil{National Radio Astronomy Observatory, Socorro, NM 87801, USA}

\author{M. Rupen}
\affil{National Radio Astronomy Observatory, Socorro, NM 87801, USA}

\author{B. Carlson}
\affil{National Research Council of Canada, Penticton, B.C. V2A 6J9, Canada}

\begin{abstract}
\noindent We report observations of type III radio bursts at decimeter wavelengths (type IIIdm bursts) -- signatures of suprathermal electron beams propagating in the low corona -- using the new technique of radio dynamic imaging spectroscopy provided by the recently upgraded Karl G. Jansky Very Large Array (VLA). For the first time, type IIIdm bursts were imaged with high time and frequency resolution over a broad frequency band, allowing electron beam trajectories in the corona to be deduced. Together with simultaneous hard X-ray (HXR) and extreme ultraviolet (EUV) observations, we show these beams emanate from an energy release site located in the low corona at a height below $\sim\!15$ Mm, and propagate along a bundle of discrete magnetic loops upward into the corona. Our observations enable direct measurements of the plasma density along the magnetic loops, and allow us to constrain the diameter of these loops to be less than 100 km. These over-dense and ultra-thin loops reveal the fundamentally fibrous structure of the Sun's corona. The impulsive nature of the electron beams, their accessibility to different magnetic field lines, and the detailed structure of the magnetic release site revealed by the radio observations indicate that the localized energy release is highly fragmentary in time and space, supporting a bursty reconnection model that involves secondary magnetic structures for magnetic energy release and particle acceleration.

\end{abstract}


\section{Introduction}

Certain types of solar activity -- flares and jets -- are powered by the impulsive release of energy through fast magnetic reconnection. In the standard scenario, a reconnection region located in low corona \citep{1994Natur.371..495M, 2003ApJ...596L.251S, 2006A&A...456..751B} accelerates particles to nonthermal energies. These particles carry a large fraction of the released energy and play an important role in energy transport processes \citep{1976SoPh...50..153L}. Electron beams can form through the propagation process, if not from the acceleration mechanism itself \citep{1998ARA&A..36..131B}, and travel upward and downward in the corona along magnetic field lines. The downward propagating beams will eventually be stopped by the dense chromospheric plasma and produce hard X-ray (HXR) emission. The chromospheric plasma is heated by these beams to high temperatures and expands hydrodynamically into the corona where it is observed in EUV and/or soft X-ray (SXR) wavelengths \citep{2008LRSP....5....1B}. However, direct information about these fast electron beams and the region from whence they originate has remained elusive. 

One method to probe these beams is through radio observations. The radio signature of coronal electron beams, first discovered in the late-1940s at meter wavelengths \citep{1950AuSRA...3..387W}, is called a type III radio burst. It is the result of the nonlinear conversion of Langmuir waves generated by beam instabilities to electromagnetic radiation at the fundamental or harmonic of the local electron plasma frequency, $f_{pe}\approx 9\sqrt{n_e}$ kHz, where $n_e$ is the electron number density. Since $n_e$ varies with height in the corona, the radio emission from the electron beam drifts from high to low frequencies for upward-propagating beams, and in the opposite sense for downward-propagating beams. Their high frequency counterpart -- type IIIdm bursts -- was first observed in the early 1960s \citep{1961ApJ...133..243Y}, and was subsequently found to be closely associated with impulsive energy release in flares \citep{1981ApJ...247.1113K, 1984SoPh...90..383D, 1985SoPh...97..159A, 2011SSRv..159..225W}. For these reasons, type IIIdm bursts are believed to be an important diagnostic of impulsive magnetic energy release. 

A barrier to exploiting type IIIdm emissions has been the lack of true ``radio dynamic imaging spectroscopy'', where simultaneous imaging observations are available at each frequency and time that a burst is observed. Imaging observations of type III, type IIIdm, or type U bursts (a type III in a closed magnetic loop) have been reported at one, or a few, discrete frequencies \citep[e.g.,][]{1992ApJ...391..380A, 1996A&A...306..299R, 1997A&AS..123..279A, 2001A&A...371..333P}, but radio dynamic imaging spectroscopy has not been possible. Here, we report the first use of radio dynamic imaging spectroscopy with the upgraded VLA to observe type IIIdm bursts during a coronal jet event that accompanied a solar flare. In \S2 we discuss the observational setup of the VLA and the data. We also present observations from a number of complementary instruments. In \S3 we discuss the implications of these data and place them in an interpretive context. We briefly conclude in \S4. 

\section{Observations}

First dedicated in 1980, the VLA has been recently upgraded with state-of-art receivers and electronics which have greatly increased its capabilities \citep{2011ApJ...739L...1P}. While the VLA is a general-purpose radio telescope, provisions were made to enable solar observing. Particularly important are the relatively large instantaneous bandwidths for imaging, the large number of spectral channels, and the fast sampling times available, which enable the new observing technique of radio dynamic imaging spectroscopy. 

The first solar observations by the upgraded VLA were made on 2011 November 5 when the VLA was in the D configuration, for which the longest antenna baseline is 1 km. The observations were made in the 1-2 GHz frequency band ($\lambda=15-30$ cm) using 1024 spectral channels, each 1~MHz in bandwidth, with a time resolution of 100 ms. Due to limitations in data throughput at that time, only 17 antennas could be used. These were nevertheless sufficient for imaging the source of radio emission. The angular resolution of the 17-antenna array was $35''\times 95''$ at 1.5 GHz at the time of observation, which scales linearly with wavelength. Both senses of circular polarization were observed. Unfortunately, a hardware error resulted in the loss of most of the data from 1.5-2.0 GHz, rendering it unsuitable for imaging although dynamic spectroscopic information is available. We therefore focus on the data from 1.0-1.5 GHz. Calibration of the antenna gains and frequency bandpasses was performed by referencing the observations to standard sidereal sources. When observing the Sun, 20 dB attenuators were inserted in the signal path. Their presence introduces fixed perturbations to the antenna gains that were measured prior to the observations and subsequently corrected by calibration.

\begin{figure*}
\begin{center}
  \includegraphics[width=0.9\textwidth]{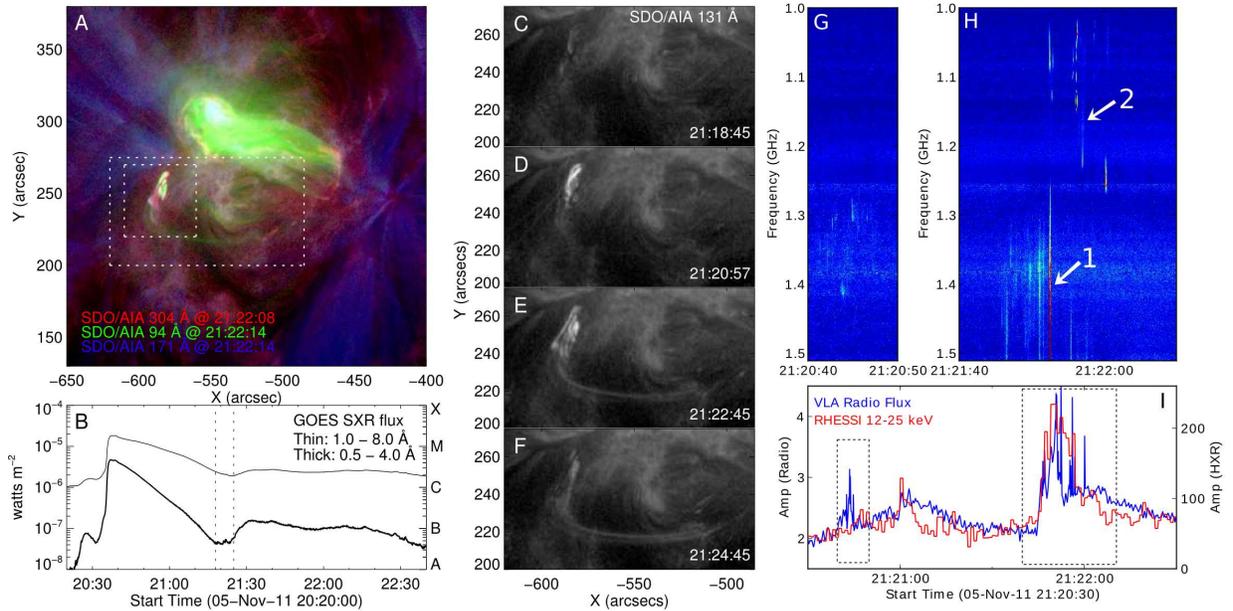}
  \caption{(A) Composite image of AR 11339 showing emission in 304 \AA\ (red, 0.1 MK), 94 \AA\ (green, 6 MK), and 171 \AA\ (blue, 0.8 MK) from SDO/AIA. (B) GOES SXR flux showing the M1.8 flare (peaks around 20:35 UT) and its aftermath. The coronal jet occurred between 21:18-21:25 UT (dashed vertical lines). (C-F) Dynamic evolution of the coronal jet and the subsequent loop brighting in AIA 131 \AA\ (0.4 MK and 10 MK) images. The field of view (FOV) is indicated by the large box in Fig. 1A. (G-H) VLA dynamic spectra in which type IIIdm bursts (the bright vertical or nearly-vertical features) are present. (I) VLA radio (blue) and RHESSI 12-25 keV HXR (red) light curves from 21:20:30-21:22:30 UT, showing the temporal correlation of the radio flux and type IIIdm bursts with the HXR flux. The small and large boxes correspond to the time ranges of the dynamic spectra shown in (G) and (H).}\label{fig:fig1}
\end{center}
\end{figure*}

\begin{figure}
\begin{center}
  \includegraphics[width=0.48\textwidth]{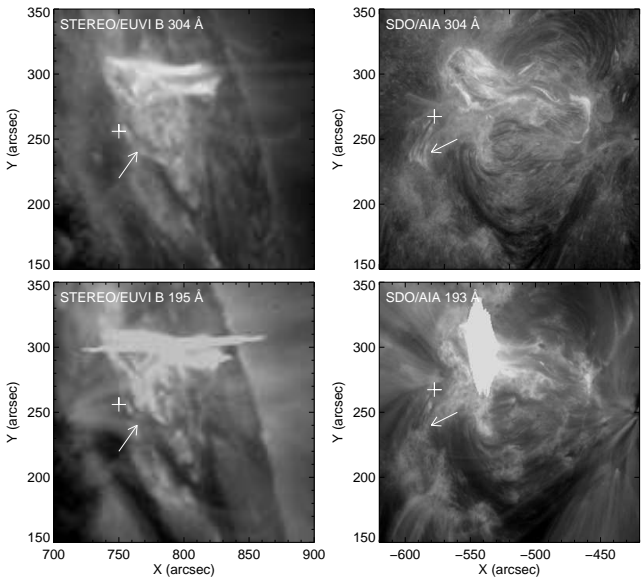}
  \caption{The coronal jet (marked by the white arrows) seen by both STEREO/EUVI B (left column) and SDO/AIA (right column) at around 21:26 UT. SDO was viewing AR 11339 from the Earth, while the STEREO B satellite was trailing the Earth by 103$^{\circ}$ at the time of the observation thereby viewing AR 11339 from the east. X- and Y-axes are aligned with solar east-west and south-north respectively in each perspective. White crosses mark the same location at the foot of the EUV jet, based on direct coordinate transformations.}\label{fig:fig2}
\end{center}
\end{figure}

Type IIIdm bursts were observed in association with a coronal jet during the aftermath of a GOES class M1.8 SXR flare (Fig. 1B) in active region (AR) 11339. The cross-power dynamic spectra are shown in Fig. 1G-H, where each bright vertical or nearly-vertical feature represents an individual burst. The largest group of bursts occurred between 21:21:40-21:22:10 UT (Fig. 1H), well correlated with a peak in the 12-25 keV HXR emission observed by the Ramaty High Energy Solar Spectroscopic Imager (RHESSI) \citep{2002SoPh..210....3L} (Fig. 1I, the red curve), indicating the release of nonthermal electrons. At 21:21:00 UT, another (weaker) HXR peak can be seen, during which the radio continuum has a small enhancement. A small cluster of narrow-band type III-like bursts can be found around the time of this HXR peak (Fig. 1G). The coronal jet was observed by the Solar Dynamics Observatory (SDO) Atmospheric Imaging Assembly (AIA) \citep{2012SoPh..275...17L} (Fig. 1A) during this energy release event, followed by subsequent brightening of closed loops (Fig. 1C-F). The jet was also observed by the EUV Imager (EUVI) aboard STEREO B \citep{2004SPIE.5171..111W} that observed the Sun from the east (Fig. 2). Interestingly, no concurrent interplanetary type III bursts were detected by the STEREO/WAVES \citep{2008SSRv..136..487B}. 

\begin{figure*}
\begin{center}
  \includegraphics[width=0.9\textwidth]{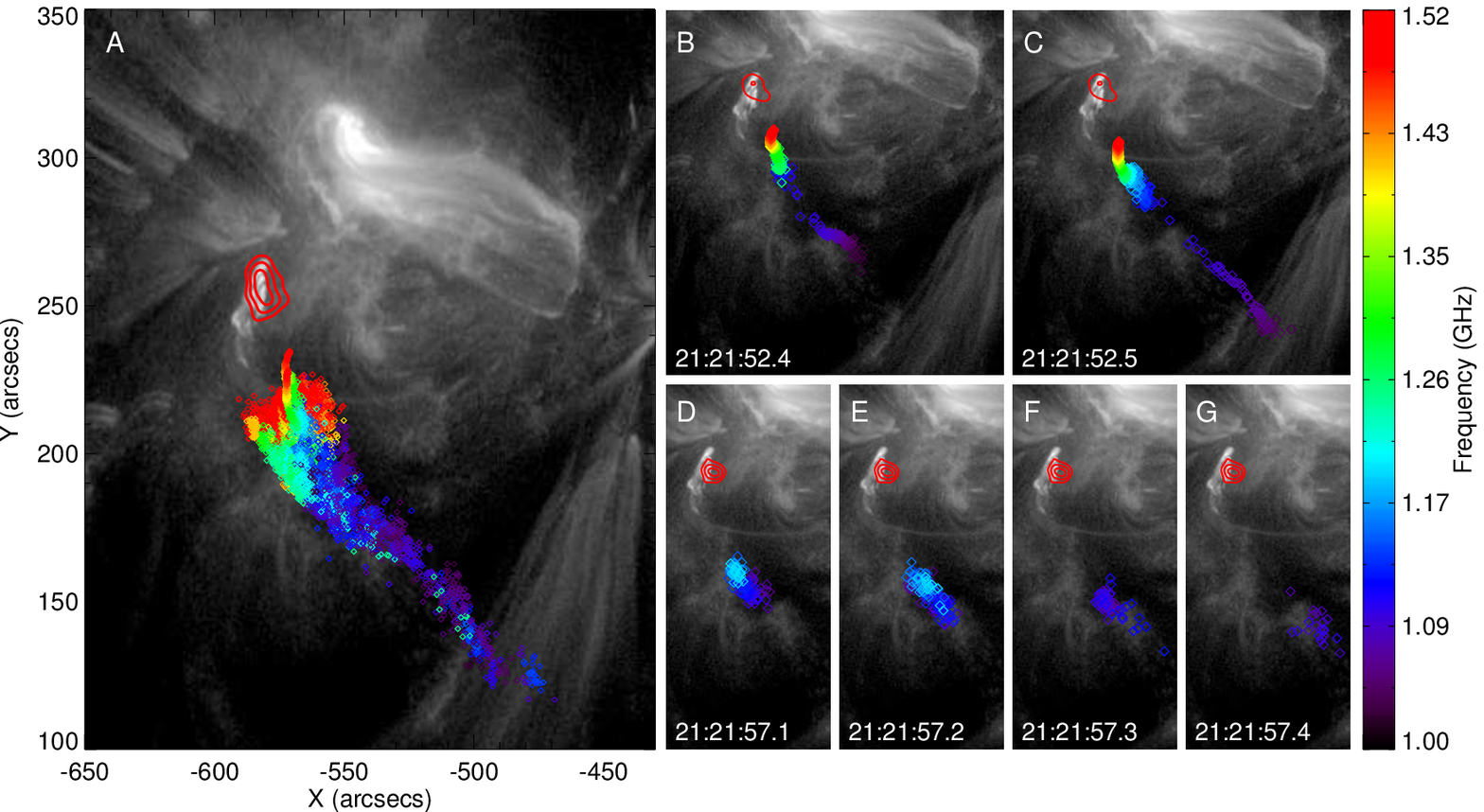}
  \caption{(A) Emission centroids of all type IIIdm bursts observed from 21:20:30-21:22:10 UT, colored from blue to red in increasing frequencies, showing electron beam trajectories in projection. Background is the SDO/AIA 131 \AA\ image at 21:22:09 UT. Red contours are the 12-s integrated 12-25 keV HXR emission during the second HXR peak (around 21:21:50 UT). Emission centroids of a temporally resolved type IIIdm burst observed from 21:21:52.4-21:21:52.6 UT (shown by arrow ``1'' in Fig. 1H) are shown by (B-C) for two successive 100-ms integrations. Another temporally-resolved type IIIdm burst from 21:21:57.1-21:21:57.5 UT (arrow ``2'' in Fig. 1H) is shown by (D-G) for four successive 100-ms integrations. Red contours are the 12-25 keV HXR source with a 4-s integration closest to the type-IIIdm-burst times. }\label{fig:fig3}
\end{center}
\end{figure*}

Dynamic imaging spectroscopy with the VLA allows each pixel in the dynamic spectrum to be imaged. Although the type IIIdm bursts themselves were unresolved by the VLA, the radio source centroid position can be determined with high accuracy. The high signal-to-noise ratio of these bursts allows their positions to be fit to $\sim\!1$\% of the nominal resolution, or $\lesssim\!1$ Mm. Examples of the fitted centroid locations for two temporally-resolved bursts (indicated by the arrows ``1'' and ``2'' in Fig. 1H) are shown in Fig. 3B-C and Fig. 3D-G, colored from blue to red corresponding to increasing frequency. At any given time, the centroid locations at different frequencies follow well-defined trajectories, distributed from southwest (bottom-right) to northeast (upper-left) with increasing frequencies. All the trajectories fall within an envelope with the high frequency end (high density, thus low coronal heights) originating near the location of the EUV jet and the HXR sources low in the atmosphere (Fig. 3A, showing a superposition of all the type-IIIdm centroid locations from 21:20:30-21:22:10 UT). In addition, for bursts that are temporally-resolved, the emission drifts from high to low frequencies with a rate of 0.3-1 GHz/s, and their centroid locations move to the southwest direction with time, indicating that the electron beams were propagating upward away from the EUV and HXR source (e.g., Fig. 3D-G). These observations show the electron beam trajectories directly and, hence, the coronal magnetic field lines along which they streamed away from the reconnection site.

\section{Results and Discussion}

The close spatiotemporal association of the type-IIIdm-burst trajectories and the HXR footpoints (Fig. 3B-G) suggests that the X-ray-producing downward-propagating electron beams and the type-IIIdm-emitting upward-propagating electron beams originate from a common energy release site, confirming previous ideas based largely on the temporal correlation between HXR/type-IIIdm emission \citep{1984SoPh...90..383D, 1995ApJ...455..347A}. It also places the reconnection site at a location above the HXR footpoints and below the height of the highest-frequency type-IIIdm-burst sources; the latter is estimated to be $\approx\!15$ Mm as we discuss further below. This energy release event was likely associated with evolving satellite magnetic polarities in AR 11339 (Fig. 4A-D). They coincided spatially with the HXR footpoint source (Fig. 4A-D) and the base of the EUV jet (Fig. 4E-H). The evolution of the HXR source and the EUV jet suggests that energy release and plasma heating took place in two stages: first, a HXR footpoint source appeared over the northern satellite polarity (Fig. 4A) during the first HXR enhancement around 21:21:00 UT, while an EUV jet was initiated (Fig. 4F; the timing was estimated within the 12-s cadence of the SDO/AIA images) that involved eruption of an arcade-shaped feature. During the second HXR enhancement peaked around 21:21:50 UT, the HXR footpoint source moved gradually to the southern satellite polarity (Fig. 4B-D), while another eruptive EUV jet was initiated (Fig. 4G-H). These multi-wavelength observations allow us to propose a self-consistent scenario describing the physical processes that occurred during this event (Fig. 5). First, interaction of the satellite magnetic polarities with the surrounding magnetic flux triggered magnetic reconnection above the northern neutral line around the first HXR peak (indicated by the ``X'' symbol). Both upward- and downward-propagating electron beams were produced, resulting in the observed type IIIdm bursts and HXR footpoint emission (the shaded area). The released magnetic energy can trigger instabilities and initiate ``blowout'' EUV jets that involve eruption of a magnetic arcade \citep{2010ApJ...720..757M}. Magnetic field lines were stretched and led to further reconnections above the southern magnetic neutral line as well. The accelerated particles led to subsequent brightenings of the southern EUV loops, HXR footpoints and type IIIdm bursts. 

\begin{figure*}
\begin{center}
  \includegraphics[width=0.9\textwidth]{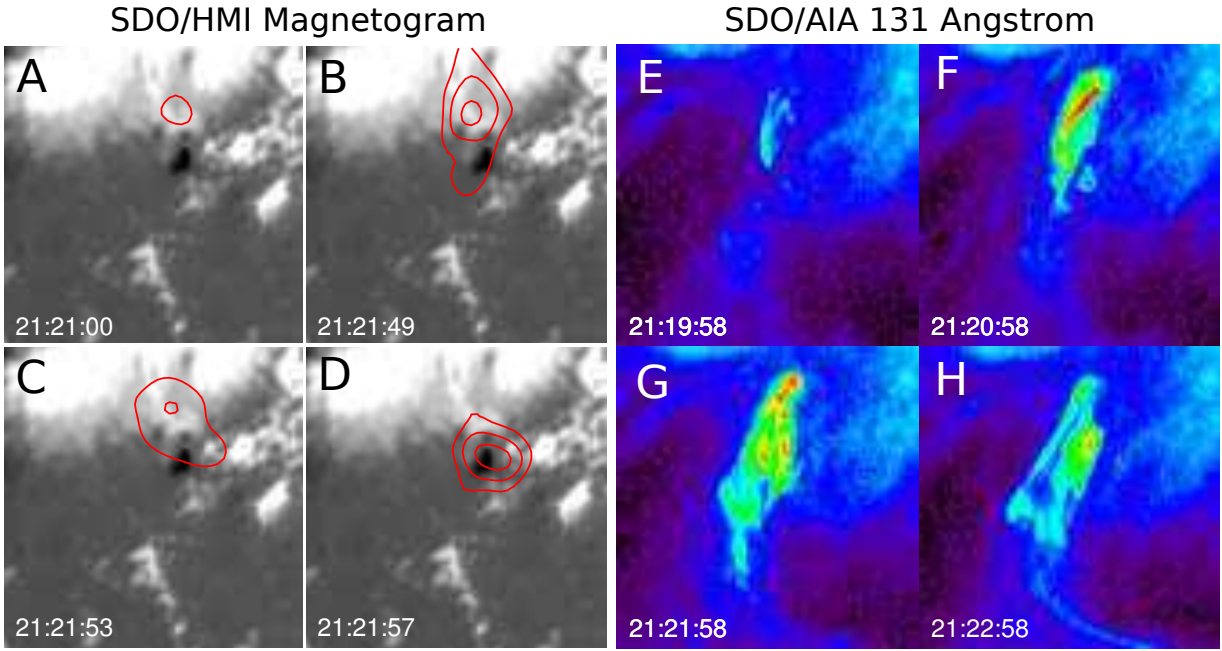}
  \caption{(A-D): A series of 12-25 keV HXR emission (red contours) overlaid on the SDO Helioseismic and Magnetic Imager (HMI) \citep{2012SoPh..275..207S} magnetogram observed during the two HXR peaks around 21:21:00 UT (A) and 21:21:50 UT (B-D). (E-H): A detailed view of the coronal jet showing its temporal evolution in SDO/AIA 131 \AA\ images. The FOV of the images is indicated by the small box in Fig. 1A.}\label{fig:fig4}
\end{center}
\end{figure*}

\begin{figure}
\begin{center}
  \includegraphics[width=0.48\textwidth]{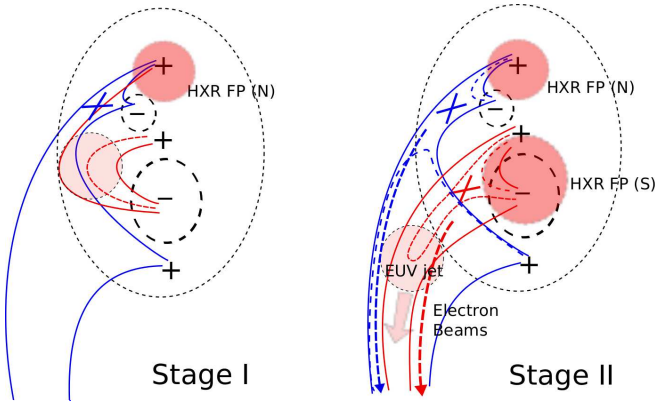}
  \caption{Schematic illustration of the magnetic field evolution, reconnection, EUV jet, and generation of the electron beams and HXR footpoint sources during the impulsive energy release event. See the text for descriptions.}\label{fig:fig5}
\end{center}
\end{figure}

The VLA data can be used to derive the electron number density $n_e$ along the type-IIIdm-emitting loops since it is directly related to the plasma frequency or its harmonic. We assume that the emission is likely harmonic plasma radiation for two reasons. First, fundamental plasma radiation tends to be more highly circularly-polarized than its harmonic \citep{1980A&A....88..203D}. Yet the majority of the observed type IIIdm bursts is weakly polarized ($\lesssim\!20$\%). Second, fundamental plasma radiation results in a group delay of $\gtrsim\!0.5$ s in its transit to Earth compared to HXR photons, while the delay is expected to be far less for harmonic radiation \citep{1984SoPh...90..383D}. No such delay $\gtrsim\!0.5$ s is found when the HXR and radio data are cross-correlated. Therefore, we have $n_e\approx (f/18\ \rm{kHz})^2=3.3-7\times 10^9$ cm$^{-3}$ over the frequency of 1.0-1.5 GHz, distributed over a height range of 30-80 Mm for the majority of the type IIIdm source centroids. The radio source height was obtained by measuring the projected distance from the HXR footpoints along the beam trajectories. Based on an estimate of the inclination angle of the EUV jet ($\sim\!45^{\circ}$) from the simultaneous SDO/AIA and STEREO/EUVI B stereoscopic observations (Fig. 2), and assuming a similar inclination for the type IIIdm bursts, the de-projected height is obtained. Fitting the ensemble of type IIIdm trajectories to a simple hydrostatic density model yields a best-fit density scale height $L_n=n_e(-dn_e/dh)^{-1}\approx 40$ Mm, corresponding to a temperature of 0.8 MK assuming hydrostatic equilibrium. Returning to the location of the reconnection site, the height of the highest-frequency type-IIIdm-burst sources is found to be as low as $\sim\!20$ Mm from the imaging results. Considering that some bursts occurred at frequencies as high as 2 GHz, for which imaging is not available, the maximum height of the reconnection site can be reduced to $\sim\!15$ Mm within the context of this density model. A striking feature of the type-IIIdm-source distributions is their frequency variation across the envelope of trajectories (Fig. 3A); that is, for a given frequency the emission is not necessarily located at the same height from burst to burst, suggesting that the type-IIIdm-emitting loops are inhomogeneous perpendicular to the magnetic field, with $\Delta n_e/\langle n_e\rangle \approx 25\%$. For the limited number of temporally-resolved bursts the electron beam velocity $v_b$ can be directly measured from the movement of the type IIIdm source centroids in time (e.g., Fig. 3D-G), which gives $v_b\approx\!0.3c$, where $c$ is the speed of light. This is consistent with the electron energy responsible for the observed HXR emission in the energy range 12-25 keV, further demonstrating that their parent electrons are accelerated from a common site.

Surprisingly, no trace of loop-like structures can be found along the type-IIIdm-burst trajectories against the background in any SDO/AIA EUV filters \citep[cf.][]{1996A&A...306..299R}, implying that the column emission measure $\xi$ of the type-IIIdm-emitting loops is too small to result in detectable emission or absorption relative to the background. To quantify this, we evaluate the observed intensity of an AIA filter image $I(x,y)=\xi (T_0,x,y)R(T_0)\Delta t$, where $\xi (T_0,x,y)$ is the emission measure assuming an isothermal plasma temperature $T_0$, $R(T_0)$ is the filter response function at $T_0$ and $\Delta t$ is the exposure time. With the estimated temperature (0.8 MK) of the type-IIIdm-emitting loops, the most probable EUV band to detect any signature is the AIA 171 \AA\ band. A negative result means the intensity contribution $\Delta I(x,y)$ by these loops is smaller than the rms intensity variation $\sigma_I$.  In regions where the type IIIdm source is present we measure $\sigma_I \sim 40-100$ DN pixel$^{-1}$ for AIA 171 \AA\ images with $\Delta t=2$ s. With $R_{171}\approx 10^{-24}$ DN cm$^5$ s$^{-1}$ pixel$^{-1}$ at 0.8 MK \citep{2012SoPh..275...17L}, the upper limit of the emission measure of these loops is estimated to be $\xi_{\rm max}=\sigma_I/\Delta t/R_{171}<5\times 10^{25}$ cm$^{-5}$. With $\xi=n_e^2 V/w^2=n_e^2 N d^2/w$, where $V$ is the total volume of $N$ identical loops along the line of sight within the instrument resolution element of size $w$ (~0.4 Mm for AIA), the upper limit of the loop diameter is $d_{\rm max}=(\xi_{\rm max} w/n_e^2)^{1/2}/N^{1/2}=l/N^{1/2}$. For a density of $n_e=3.3-7\times 10^9$ cm$^{-3}$ inside the loops, $l$ is found to be only a few $\times\!10$ to $\approx\!100$ km under rather generous assumptions, and $d_{\rm max}$ would be even smaller with a filling factor $1/N<1$.

The background plasma density and temperature can be estimated by utilizing the observed background intensities $I_b(x,y)$ from all the AIA EUV bands sensitive to a wide range of coronal temperature. We use \citet{2011SoPh..tmp..384A}'s automated forward-fitting technique to obtain a differential emission measure distribution, $d\xi_b(T,x,y)/dT$, which yields an integrated emission measure of $\xi_b\approx 5\times 10^{27}$ cm$^{-5}$ peaking at $\sim\!2.8$ MK. With $\xi_b=\int n_e(h)^2 dh$, the background density is found to vary from $6.9\times 10^8$ to $4.8\times 10^8$ cm$^{-3}$ over 30-80 Mm at which the type IIIdm bursts occur. Hence the type-IIIdm-emitting loops may be cooler (by a factor of $\sim\!3$) and denser (by roughly an order of magnitude) relative to the background coronal medium. This is consistent with the scenario proposed by \citet{1992SoPh..141..335B} for allowing the type IIIdm emission to escape. The over-density is also consistent with past conclusions based on the observed contrast between $\xi$ toward bright EUV/SXR loops and $\xi_b$ toward the background corona \citep{1992PASJ...44L.135H, 2001ApJ...560.1035A}, which suggests that the relevant active region loops may have already been supplied with high-density chromospheric plasmas through previous heating processes.  

The existence of ultra-fine magnetic structures in the corona has been previously suspected \citep[see, e.g.][and references therein]{2010LRSP....7....5R} but no current instrumentation can resolve such structures at coronal temperatures. Recent observations of ultra-fine magnetic structures have been of plasma at chromospheric temperatures \citep{2012ApJ...750L..25J, 2012ApJ...745..152A}, reporting loop diameters ranging from $\sim\!100$ to a few $\times\!100$ km. Our observations suggest the corona is ``fibrous'' in nature, consisting of many unresolved ``strands''. It should also be noted that the type-IIIdm-emitting strands have a direct connectivity to the reconnection region. Thus these strands should be even thinner down at the reconnection region with a diameter of at most 10s of km due to their general expansion with increasing coronal height \citep{2007ApJ...661..532D}. Furthermore, the multitudes of discrete electrons beams observed have access to spatially distinct coronal magnetic strands in $\lesssim\!1$ s (Fig. 3A), which indicates that the reconnection region likely consists of a large number of discrete reconnection sites in a localized spatial volume.  Our observations rather directly suggest a bursty reconnection scenario involving a localized reconnection region containing a distribution of many small-scale dynamically-evolving structures for magnetic energy release and particle acceleration. This picture is qualitatively consistent with previous work on fragmentary energy release on the Sun \citep[see, e.g.][and references therein]{1994Kluwer}, as well as more recent developments in the physics of magnetic reconnection, such as experiments in the laboratory \citep{2012Natur.482..379M} and numerical simulations \citep{2000A&A...360..715K, 2006Natur.443..553D,2007A&A...464..735K}, which show the reconnection region consists of many ``magnetic islands'' or ``filaments'' developed through magnetohydrodynamic instabilities such as tearing or kink instability. The constraints in the temporal and spatial scales of fragmentations we report here and in future observations of this kind may help guide theoretical reconnection models as well as constrain parametric inputs to numerical simulations.

\section{Conclusions}

Dynamic imaging spectroscopy of type IIIdm bursts with the VLA has allowed us to map the trajectories of electron beams produced by magnetic energy release during a coronal jet that occurred in the aftermath of a flare. Electrons escaped along discrete, ultra-fine strands into the upper atmosphere, producing the observed type IIIdm bursts. Together with magnetic, EUV, and HXR data, we showed that the energy release site was associated with a coronal jet and that the magnetic reconnection process involved different locations at different times and that, furthermore, the density at each location varied significantly. The spatial scales in the reconnection sites are likely 10s of km or less.  We conclude that the magnetic energy release process is highly fragmentary and that the surrounding coronal medium is fibrous in nature. 

\acknowledgements
The National Radio Astronomy Observatory is a facility of the National Science Foundation operated under cooperative agreement by Associated Universities, Inc. The authors wish to thank the teams of SDO/AIA, SDO/HMI, STEREO/EUVI, and RHESSI for providing the relevant data. BC acknowledges support by NSF grant AGS-1010652 to the University of Virginia and support from the NRAO through the Resident Shared Risk Observing program.

\end{document}